\documentclass[12pt,preprint]{aastex}

\newcommand{\msunyr}{M$_\odot$ \thinspace yr$^{-1}$}
\newcommand{\msunpc}{M$_\odot$ pc$^{-2}$}
\newcommand{\msunyrkpc}{M$_\odot$ pc$^{-2}$ Myr$^{-1}$}
\newcommand{\Halpha}{H$\alpha$}

\newcommand{\kms}{km s$^{-1}$}
\newcommand{\msun}{$M_{\sun}$}

\begin{document}

\title{High star formation rates in turbulent atomic-dominated gas in the interacting galaxies IC 2163 and NGC 2207}

\author{Bruce G. Elmegreen\altaffilmark{1},
Michele Kaufman\altaffilmark{2}, Fr\'ed\'eric Bournaud\altaffilmark{3}, Debra Meloy
Elmegreen\altaffilmark{4}, Curtis Struck\altaffilmark{5}, Elias Brinks\altaffilmark{6},
Stephanie Juneau\altaffilmark{3}}

\altaffiltext{1}{IBM Research Division, T.J. Watson Research Center, P.O. Box 218, Yorktown
Heights, NY 10598; bge@watson.ibm.com}

\altaffiltext{2}{110 Westchester Rd, Newton, MA 02458, USA; kaufmanrallis@icloud.com}

\altaffiltext{3}{Laboratoire AIM-Paris-Saclay, CEA/DSM-CNRS-Universit\'e Paris Diderot,
Irfu/Service d'Astrophysique, CEA Saclay, Orme des Merisiers, F-91191 Gif sur Yvette,
France; frederic.bournaud@gmail.com; stephanie.juneau@cea.fr}

\altaffiltext{4}{Department of Physics \& Astronomy, Vassar College, Poughkeepsie, NY
12604; elmegreen@vassar.edu}

\altaffiltext{5}{{Department of Physics \& Astronomy, Iowa State University, Ames, IA
50011; struck@iastate.edu} }

\altaffiltext{6}{{University of Hertfordshire, Centre for Astrophysics Research, College
Lane, Hatfield AL10~~9AB, United Kingdom; e.brinks@herts.ac.uk} }

\begin{abstract}
CO observations of the interacting galaxies IC 2163 and NGC 2207 are combined with HI,
H$\alpha$ and 24 $\mu$m to study the star formation rate (SFR) surface density as a
function of the gas surface density. More than half of the high SFR regions are HI
dominated. When compared to other galaxies, these HI-dominated regions have excess SFRs
relative to their molecular gas surface densities but normal SFRs relative to their total
gas surface densities. The HI-dominated regions are mostly located in the outer part of NGC
2207, where the HI velocity dispersion is high, 40 - 50 km s$^{-1}$. We suggest that the
star-forming clouds in these regions have envelopes at lower densities than normal, making
them predominantly atomic, and cores at higher densities than normal because of the high
turbulent Mach numbers. This is consistent with theoretical predictions of a flattening in
the density probability distribution function for compressive, high Mach number turbulence.
\end{abstract}


\section{Introduction}
\label{intro}

The galaxies IC 2163 and NGC 2207 are undergoing a grazing collision with perigalacticon
$\sim240$ Myr ago \citep{struck05}. The tidal force exerted on IC 2163 by NGC 2207 is
prograde and nearly in-plane.  It produces an intrinsically ocular structure (a cuspy oval)
at mid-radius in IC 2163 and two tidal arms. A narrow ridge along the ocular appears to be
a galactic-scale shock front with intense star formation, several super-star clusters
including one measuring $2\times10^5\;M_\odot$ \citep{elmegreen01}, and an IR-to-radio
continuum ratio ($S$[8 \micron]/$S[6\; {\rm cm}]$) that is a factor of 2 higher than
elsewhere in the galaxies \citep{kaufman12}. HI observations \citep{elmegreen95a} reveal
photometric and kinematic major axes in IC 2163 that are nearly orthogonal, indicating high
speed streaming motions. The tidal force exerted on NGC 2207 by IC 2163 is mostly
perpendicular to NGC 2207, causing a velocity perturbation in NGC 2207 that is $\sim 200$
km s$^{-1}$ in places, and suggesting a warp with 15 kpc vertical distortion
\citep{elmegreen95b}.

Both galaxies have large HI velocity dispersions of $\sim30-50$ km s$^{-1}$, and several
large HI cloud complexes with masses of $10^8\;M_\odot$ or more
\citep{elmegreen93,elmegreen95a}. IC 2163 has an HI tidal bridge behind the eastern half of
NGC 2207, and NGC 2207 has an HI spiral arm (visible also optically) in front of IC 2163
\citep[see Figures in][where the HI contours are overlaid on an HST image]{elmegreen00}.

NGC 2207 contains a peculiar region, {\em Feature i}, called a mini-starburst by
\cite{kaufman12}. This is the brightest source for either galaxy in radio continuum, 8
\micron, 24 \micron\ and $H\alpha$; it accounts for one-quarter of the total 24 \micron\
flux for both galaxies combined. At $\lambda$ 6 cm, the radio continuum luminosity of {\em
Feature i} is 13\% of the radio continuum luminosity of the central starburst in M82. Soft
X-ray emission was found by \cite{mineo14}.

We observed NGC 2207/IC 2163 in CO(1-0) with the Atacama Large Millimeter Array (ALMA). The
present paper is devoted to determining the spatially-resolved Kennicutt-Schmidt
\citep{kennicutt12} relation in this galaxy pair.  We are interested in whether this
relation on a scale of $2.4$ kpc diameter differs from that in other galaxies, and whether
it is the same in both galaxies here. We combine the ALMA CO observations with our previous
HI, 24 \micron, and $H\alpha$ data to determine the star formation rate (SFR) surface
density, $\Sigma_{\rm SFR}$, as a function of atomic, molecular, and total gas surface
densities. Other properties of these CO observations will be discussed in Kaufman et al.
(2016).

In what follows, Section 2 describes the observations, Section 3 gives the luminosities,
SFRs, and surface densities of the selected regions, Section 4 shows the relations between
$\Sigma_{\rm SFR}$ and the gas surface density, and Section 5 discusses the implications.

\section{Observations}

IC 2163 and NGC 2207 were observed in 34 pointings using ALMA at the $^{12}$CO(1-0) frequency
of 115.27 GHz ($\lambda$ 2.6 mm). We made naturally weighted maps of CO emission with a point
spread function of $2.00^{\prime\prime} \times 1.52^{\prime\prime}$ (HPBW), beam position
angle $68.5^\circ$, and channel width $10$ km s$^{-1}$. The rms noise per channel is $3.7$
mJy beam$^{-1}$ and a CO surface brightness of 1 Jy beam$^{-1}$ corresponds to $T_b = 30.7$
K. To select areas of genuine CO emission, we made a blanking mask by convolving this cube to
$6^{\prime\prime}\times6^{\prime\prime}$ HPBW, where the noise was 7.9 mJy/beam. Regions were
left unblanked if they exceeded 2.5 times this rms noise for at least 2 adjacent channels in
an area of at least 1 arc sec. This means that regions with line-of-sight $N(H_2) <
2.45\;M_\odot$ pc$^{-2}$ were blanked out. After correcting for primary beam attenuation, we
made the $551\times301$ pixel $\times64$ channel subcube used in this paper that contains all
of the unblanked CO emission. In the subcube, the product of the rms noise and the channel
width is equivalent to a line of sight $N(H_2) = 3.27\;M_\odot$ pc$^{-2}$, which is a measure
of our uncertainty. (The apertures A10, A16, A33, A38, and A39 discussed below have mean
values of line-of-sight $N(H_2)$ below this because of blanked pixels.) The total integrated
CO line flux in the subcube is $ S(CO)=505$ Jy km s$^{-1}$. For more details on the observing
procedure, see \cite{kaufman16}.

To convert CO emission into molecular mass $M(H_2)$, we use $X_{\rm CO}= 1.8 \pm 0.3
\times10^{20}$ H$_2$ cm$^{-2}$ (K \kms)$^{-1}$ from \cite{dame01} and a distance $D =
35\pm2.5$ Mpc from the NASA/IPAC Extragalactic Database (NED) for a Hubble constant of $H =
73$ km s$^{-1}$ Mpc$^{-1}$ corrected for infall towards Virgo. The scale is 170 pc per
arcsec. To convert surface density to face-on, we took inclinations of $i = 40^\circ$ for IC
2163 and $35^\circ$ for NGC 2207 \citep{elmegreen95b}.  We discuss the implications of
different $X_{\rm CO}$ and $i$ in Section \ref{sfrcorrel}.

The 24 \micron\ flux density, $S_{\nu}(24\mu m)$, of {\em Feature i} and the galaxies as a
whole were measured from the MIPS 1 pBCD image obtained from the Spitzer archive in 2013; the
resolution is $6^{\prime\prime}$. Other star-forming regions were measured on the HiRes
deconvolution image ($\sim1.9^{\prime\prime}$ resolution) by \cite{velusamy08}, who provided
their HiRes 24 \micron\ data to us as a FITS image \footnote{MIPS 1 BCD images from Spitzer
prior to April 2007 suffer from two software bugs which cause the brightness of sources as
bright as {\em Feature i} to be underestimated. We compared our Spitzer 24 \micron\ MIPS 1
pBCD (level 2) image retrieved from the Spitzer archive in 2013 with the MIPS 1 image in
\cite{elmegreen06}. We found that {\em Feature i} was the only source in NGC 2207/IC 2163
affected by the software bug, and that its flux density is a factor of 1.9 times greater than
in \cite{elmegreen06} and \cite{kaufman12}. Thus {\em Feature i} accounts for 23\%, rather
than 12\%, of the 24 \micron\ flux density from the galaxy pair. We use the corrected flux
here.}. According to Velusamy et al., the HiRes deconvolution conserves fluxes of point
sources to better than 5\%. This uncertainty is in addition to the 5\% uncertainty in the
MIPS 1 photometric calibration. The HiRes deconvolution has about the same resolution as our
CO data.  This high resolution minimizes line-of-sight corrections for diffuse emission at 24
\micron\ that is not coincident with the star-forming region \citep{leroy12}. The absolute
calibration of the 24 \micron\ term in the SFR leads to an uncertainty of about 25\% in the
absolute determination of the SFR \citep{leroy12}.

The HI is from VLA observations obtained in 1990 using a hybrid CnB array with an extended
north arm to compensate for the low latitude of the source. The duration on source was 6
hours \citep[for more details, see][]{elmegreen95a}. The map representing column density,
$N(HI)$, comes from an HI cube with a channel width of 21 km s$^{-1}$, rms noise 0.73 mJy
beam$^{-1}$, and FWHM of the point spread function equal to $13.5^{\prime\prime} \times
12^{\prime\prime}$. The product of the rms noise times the channel width is equivalent to a
line-of-sight $N(HI) = 0.84\;M_\odot$ pc$^{-2}$. Before creating the HI column density map,
the HI data cube was masked in a similar manner to the CO data described above; none of our
apertures contain blanks in HI.

The H$\alpha$ data stem from narrowband H$\alpha$ images and broadband $R$ images obtained by
Deidre Hunter with the Lowell 1.1m telescope in 1999 \citep{elmegreen01}. Four H$\alpha$
images were combined for a total time of 3600 s, and three R band images were combined for a
total of 900 s. The narrow filter excludes the redshifted [NII] line at 6583 \AA, but
includes the 6548 \AA\ line, which may contribute as much as 15\% of the light for solar
metallicity. Considering equation \ref{sfeq} below, a systematic 15\% decrease in the
H$\alpha$ flux after removal of [NII] amounts to an average decrease in $\Sigma_{\rm SFR}$ of
5\%.

Figure 1 displays the chosen regions as circular apertures overlaid on the CO integrated
intensity image, the HiRes $24\;\mu$m image, and the $H\alpha$ image, which has a point
spread function of $4.2^{\prime\prime}\times3.6^{\prime\prime}$. The apertures were selected
to have minimal overlap and to enclose regions with prominent 24 \micron, $H\alpha$, CO
and/or HI emission. They include three of the massive HI clouds in IC 2163 (I2, I3, I5) and
four in NGC 2207 (N2, N3, N4, N6), even though some have little star formation.  Figure 2
shows the HI contours on CO for IC 2163 (top) and NGC 2207 (bottom), where the separate HI
contributions from each galaxy were determined from the line kinematics. For IC 2163 the
lowest contour level is $5\;M_\odot$ pc$^{-2}$, which is equivalent to three times the rms
noise over 2 channel widths. For NGC 2207 we omitted this contour level for clarity.

Table \ref{tab-global} compares global values of $M(H_2)$, $M(HI)$, the 24 \micron\ flux
density, $S_{\nu}(24\mu m)$, and the $H\alpha$ flux, $S(H\alpha)$, for the fields displayed
in Figure 1. For the values of $M(H_2)$, $S_{\nu}(24\mu m)$, and $S(H\alpha)$, we chose a
right ascension of $06^{\rm h} 16^{\rm m} 25.60^{\rm s}$ as the dividing line between IC
2163 and NGC 2207. For $M(HI)$, we included the tidal bridge of IC 2163 behind NGC 2207 and
the spiral arm of NGC 2207 in front of IC 2163. The values listed under $M(HI)$ (total)
include the HI emission outside the field in Figure 1; the values listed under $M(HI)$ (CO
field) are only for the field in that figure. For IC 2163, the value of $S_{\nu}(24\mu m)$
measured on the HiRes image is 84\% of that measured on the MIPS 1 pBCD image (the HiRes
image is less sensitive to faint extended emission).

Table \ref{tab-global} indicates that IC 2163 has a total molecular mass $M(H_2) =
2.1\times10^9\; M_\odot$, which corresponds to an integrated CO line flux of 240 Jy km
s$^{-1}$, and NGC 2207 has $M(H_2) = 2.3\times10^9\;M_\odot$ for a CO line flux of 265 Jy
km s$^{-1}$. If we restrict the HI to the field of Figure 1, then the global ratio of
molecular to atomic gas is 4 times greater in IC 2163 than in NGC 2207. For most of the 15
apertures in IC 2163 (except those on the massive HI clouds), the column density $N(H_2$)
dominates $N(HI)$, while for most of the 29 apertures in NGC 2207, $N(HI)>N(H_2)$. Thus,
the eyelid shock in IC 2163 is currently more effective than the spiral arms of NGC 2207 in
converting HI to H$_2$. Also, the global ratio of $M(H_2)$ to 24 \micron\ flux density is
$\sim2$ times greater in IC 2163 than in NGC 2207, and the global ratio of $M(H_2)$ to
$H\alpha$ flux is $\sim3$ times greater in IC 2163 than in NGC 2207. These molecular
excesses compared to star formation in IC 2163 are not as large as the molecular excesses
compared to atomic gas as traced by HI. This difference suggests that the encounter has
converted a high fraction of HI into H$_2$ in IC 2163 but has not (yet) converted the additional
molecules into stars.

\section{Star formation rates}
\label{sect:sfr}

A combination of the continuum-subtracted $H\alpha$ image \citep{elmegreen01} and the
Spitzer MIPS 24 \micron\ image was used to obtain the SFR, following \cite{kennicutt09},
\begin{equation}
{\rm SFR}(M_\odot \;{\rm yr}^{-1}) = 5.5\times10^{-42}\left(L[H\alpha] + 0.031 L[24\mu m]\right)
\label{sfeq}
\end{equation}
where $L(H\alpha)$ and $L(24\mu m)$ are in erg s$^{-1}$, and $L(24\mu m) = \nu L_{\nu}$. We
chose source apertures with a diameter of $14^{\prime\prime} = 2.4$ kpc. This choice was
governed by the HI resolution and by the need to have a sufficiently large aperture to
avoid stochastic effects from including too few O stars when using \Halpha\ as a
star-formation tracer.

For the 24 \micron\ measurement of {\em Feature i}, we did a local background subtraction
and applied an aperture flux correction factor of 1.61 from the MIPS Instrument
Handbook\footnote{http://irsa.ipac.caltech.edu/data/SPITZER/docs/mips/mipsinstrumenthandbook/}.
It was not necessary to remove background for the CO or HI sources, nor for the other HiRes
24 \micron\ sources. For $H\alpha$, we did a global background subtraction but not a local
subtraction because there would be contamination from adjacent sources in many cases.

Foreground Galactic extinction $A_{\rm V} = 0.238$ mag from NED \citep[using][]{schlafly11}
gives an extinction at $H\alpha$ equal to $0.238\; {\rm mag}/1.28 = 0.186$ mag. In
addition, IC 2163 is affected by extinction from the foreground spiral arm of NGC 2207.
Since the outer part of NGC 2207 is likely to be metal poor, we take $A_{\rm V}{\rm (mag)}
= (0.35 \pm 0.18)\times10^{-21} N(HI)$ for this region \citep{elmegreen01}. This additional
extinction is applied to 8 of the 15 regions in IC 2163. These 8 have the highest
foreground $N(HI)$ and the largest corrections ($A_{\rm V}(H\alpha) > 0.45$ mag), with an
average $A_{\rm V}(H\alpha) = 0.85 \pm 0.4$ mag from the foreground HI in NGC 2207. We
assume that $H\alpha$ and 24 \micron\ emission from the foreground part of NGC 2207 are
negligible.

Table \ref{tab-location} compiles the locations, luminosities, SFRs, $\Sigma_{\rm SFR}$, and
surface densities for H$_2$ and HI in all of the chosen apertures. The regions associated
with massive HI clouds are indicated by the footnote {\it h}. The SFR is lower than 0.001 in
the massive HI cloud apertures A10 and A33. This is below the limit suggested by
\cite{leroy12} where the SFR can be reliably determined using the above method; the actual
values do not contribute to the conclusions of this paper.

\section{Star formation - surface density correlation}
\label{sfrcorrel}

The conventional way to consider star formation on galactic scales is in terms of the
correlation between $\Sigma_{\rm SFR}$ and gas surface density (the ``Kennicutt-Schmidt''
relation), either for HI ($\Sigma_{\rm HI}$), H$_2$ ($\Sigma_{\rm H2}$), or the sum of these
\citep{kennicutt12}.

Figure \ref{fig:n2207_alma_ks1c} shows these correlations. Figure
\ref{fig:n2207_alma_ks1c}(a) has the total gas relation, Figure
\ref{fig:n2207_alma_ks1c}(b) has H$_2$ alone, \ref{fig:n2207_alma_ks1c}(c) has HI alone,
and \ref{fig:n2207_alma_ks1c}(d) shows $\Sigma_{\rm SFR}$ versus $\Sigma_{\rm
H2}/\Sigma_{\rm HI}$, the molecular ratio. Figure \ref{fig:n2207_alma_ks1c}(d) is divided
into three parts: where $\Sigma_{\rm SFR}> 0.01$, the green points are for a molecular
ratio less then 1 and the red points are for a molecular ratio greater than 1. Where
$\Sigma_{\rm SFR}< 0.01$, the blue points are plotted for all molecular ratios. These
colors have the same meaning in the other panels.

First consider the molecular relation in Figure \ref{fig:n2207_alma_ks1c}(b). In the THINGS
survey \citep{bigiel08,leroy08}, this was linear with a constant consumption time per CO
molecule of 1-2 Gyr. This is the case here too, but only for the red and blue points, i.e.,
for the molecular-dominated gas at high SFR and all of the gas at low SFR. The HI-dominated
gas at high SFR (green points) lies high off the linear molecular relation. The scatter in
the molecular Kennicutt-Schmidt relation in \cite{bigiel08} is $\pm0.2$ dex for $\Sigma_{\rm
SFR}$, which is three times smaller than the displacement of the green points from the
red+blue point correlation in Figure \ref{fig:n2207_alma_ks1c}(b).  The H$_2$ consumption
time for the HI-dominated gas is $\sim300$ Myr, which is relatively fast compared to the time
for the H$_2$-dominated gas. The red circles around the points are for locations in the
ocular rim of IC 2163.

The total-gas relation in Figure \ref{fig:n2207_alma_ks1c}(a) has a consumption time for
total gas of about 1 Gyr in regions with high SFRs (green and red points). This time is
normal for molecular gas, but here it includes HI and also applies in HI-dominated regions
(green points). The blue points (low SFR) have longer total-gas consumption times, between
$\sim3$ and 10 Gyr.

In Figure \ref{fig:n2207_alma_ks1c}(c), the molecular-dominated regions (red points) have
relatively low HI and the HI-dominated regions (green points) have relatively high HI. The
values for the HI surface density in the HI-dominated, high SFR regions are unusually high:
$\Sigma_{\rm HI}$ is greater than $20\;M_\odot$ pc$^{-2}$ in many cases, which is twice as
large as the saturated value of HI in the THINGS survey \citep{bigiel08,leroy08}.

What is peculiar about these relationships is the HI. If we ignore the HI-rich, high SFR
regions (green points), then the molecular relation in Figure \ref{fig:n2207_alma_ks1c}(b)
is normal; the nearly constant $\Sigma_{\rm HI}$ up to $\sim10\;M_\odot$ pc$^{-2}$ for the
high-SFR regions (red points) in Figure \ref{fig:n2207_alma_ks1c}(c) is normal, and the
steady increase in $\Sigma_{\rm SFR}$ with molecular fraction in Figure
\ref{fig:n2207_alma_ks1c}(d) is normal. In these galaxies, however, there is a component of
high HI column density gas where the SFR is high also. The HI is replacing some fraction of
the molecules, giving a linear relation when the SFR is plotted against the total gas (Fig.
\ref{fig:n2207_alma_ks1c}(a)). Similarly, the HI-dominated regions lie off the normal
linear law when the SFR is plotted versus only the molecular part of the gas (Fig.
\ref{fig:n2207_alma_ks1c}(b)). We also note that the molecular ratio
$\Sigma_{H2}/\Sigma_{HI}$ increases with decreasing $L(H\alpha)/[0.031L(24\mu m)]$ (not
shown) suggesting more optical extinction in the star-forming cores when the molecular
content is high.

Figure \ref{fig:n2207_alma_ks3} shows the locations of these three types of regions on the
sky using the same color scheme as in Figure \ref{fig:n2207_alma_ks1c}. The HI-rich regions
of high $\Sigma_{\rm SFR}$ are on the periphery of the galaxy that underwent the
retrograde, perpendicular encounter (NGC 2207) and are globally coincident with the areas
of high HI velocity dispersion in that galaxy (see Elmegreen et al. 1995a). The H$_2$-rich
regions of high $\Sigma_{\rm SFR}$ are mainly in the ocular ridge of IC 2163 where in-plane
tidal forces produced a compression. The low $\Sigma_{\rm SFR}$ regions are scattered
across both galaxies.

{\em Feature i} in each panel (denoted by ``i'') has $\Sigma_{\rm SFR}$ at least a factor
of $5$ higher than in any other region. The molecular mass of {\em Feature i} is $8\times
10^7$ \msun, which is not unusual for these 2.4 kpc diameter apertures. In IC 2163, 11 of
the 15 apertures have molecular masses greater than this (Table \ref{tab-location}), and
the mass in region A9 is 2.5 times greater. However, because of the high SFR, the molecular
gas consumption time for {\em Feature i}, $\Sigma_{\rm H2}/\Sigma_{\rm SFR}=50$ Myr, is
much shorter than for the other regions, and the total gas consumption time, $(\Sigma_{\rm
HI}+\Sigma_{\rm H2})/\Sigma_{\rm SFR}=118$ Myr, is shorter too, as indicated by the
positions of {\em Feature i} relative to the dashed red lines in Figures
\ref{fig:n2207_alma_ks1c}(b) and (a), respectively.

A lower $X_{\rm CO}$ as suggested for some interacting galaxies and in \cite{bournaud15}
would decrease the H$_2$ mass and strengthen the conclusion that there are HI-rich
star-forming regions with peculiar cloud structure. A higher $X_{\rm CO}$ would bring the
H$_2$ content of these regions in line with the CO-bright regions, but then $X_{\rm CO}$
would have to vary from cloud to cloud and be high primarily in the outer parts of NGC 2207,
where the peculiar HI gas is. These outer parts are not so remote that low metallicities and
an associated high $X_{\rm CO}$ are expected. Higher $X_{\rm CO}$ is also unlikely because
the high velocity dispersion should lower the CO opacity by broadening the line, producing
more CO emission per unit H$_2$ molecule \citep{bournaud15}. Inclination effects would not
seem to be causing the peculiar HI either.  The suspected warp in NGC 2207 occurs about where
this HI is, but correcting for that by lengthening the line-of-sight for gas emission only
strengthens the conclusions because it lowers the average density for the observed column of
HI and makes star formation slower, when in fact it is faster than normal.  Changing the
inclination slides points parallel to the lines of constant gas consumption time in the
Kennicutt-Schmidt relation because it affects the deprojected areas used for the ordinate and
abscissa equally.

A second version of Figure \ref{fig:n2207_alma_ks3} was made (not shown) with all of the
images convolved to $14^{\prime\prime} \times 14^{\prime\prime}$ resolution before the
measurements were made inside the same apertures. This change had the effect of decreasing
the fluxes used for the SFRs and $\Sigma_{\rm H2}$ by factors between 0.5 and 0.9, and that
moved the points in the figure slightly down and to the left in a direction nearly parallel
to the lines of constant consumption time. Considering the wings of the Gaussian, this is the
expected reduction factor for a small source located somewhere in the aperture when the
aperture diameter is the same as the FWHM of the point spread function in the convolved
image. None of the conclusions of this paper were affected by this change.

\section{Discussion}

HI-dominated regions with high SFRs are unusual and could be related to the high turbulent
speeds in this interacting pair.  The mean HI velocity dispersion for the 16 green circles in
N2207 (Fig. \ref{fig:n2207_alma_ks3}) is 42 km s$^{-1}$, compared to $\sim10$ km s$^{-1}$ in
normal galaxies \citep{tamburro09}. An interaction perturbs the smooth circular flow of gas
and causes deflected streams to intersect each other at high speeds, producing shocks and a
turbulent cascade. Numerical simulations show this effect
\citep{wetzstein07,bournaud11a,powell13}. Some turbulent energy in NGC 2207 may also come
from excited vertical motions. High velocity dispersions in interacting systems were also
reported by \cite{kaufman99}, \cite{rich15} and others.  In comparison, the THINGS galaxies
and most others used for the conventional SFR-gas correlations are not interacting.

Stronger shocks imply a greater compressive component to the turbulence, as opposed to a
normally dominant rotational component \citep{federrath08,federrath10}. Tidal forces can
increase the compressive mode too \citep{renaud14}. As a result, the probability
distribution function for density flattens from an approximately log-normal \citep{nolan15}
shape to a broad distribution where a large amount of low-density atomic gas co-exists with
a large amount of high-density star-forming gas
\citep{federrath13,federrath13b,bournaud11b,renaud12}. The star formation rate may scale
with the mass of the high density gas as usual \citep[e.g.,][]{evans14,clark14}.
Physically, this implies a change in the structure of clouds toward more extended HI
envelopes around denser H$_2$ and CO cores.

The molecular envelop traced by CO in a self-gravitating cloud usually contributes to the gas
surface density in the molecular star formation law.  Here, the envelopes may be puffed up
with high velocity dispersions, giving them low densities and a transparency that keeps them
predominantly atomic.  Then the envelopes act physically like the CO envelopes of
star-forming clouds in non-interacting galaxies, but produce a star formation law with a
significant amount of HI and dark $H_2$ substituted for CO-bright gas.

Another possibility for the HI-rich regions of high SFR is that there is large-scale
synchronization of star formation in NGC 2207, starting at perigalacticon $\sim240$ Myr ago,
whereby many of the giant molecular clouds formed in the outer parts have just now dispersed
into HI before their OB associations have significantly faded.

\acknowledgments

We are grateful to Dr. Kartik Sheth for his generous help at all stages of this project. This
paper makes use of the following ALMA data: ADS/JAO.ALMA\#2012.1.00357.S. ALMA is a
partnership of ESO (representing its member states), NSF (USA) and NINS (Japan), together
with NRC (Canada) and NSC and ASIAA (Taiwan) and KASI (Republic of Korea), in cooperation
with the Republic of Chile. The Joint ALMA Observatory is operated by ESO, AUI/NRAO and NAOJ.
The National Radio Astronomy Observatory is a facility of the National Science Foundation
operated under cooperative agreement by Associated Universities, Inc. This research has made
use of the NASA/IPAC Extragalactic Database (NED) which is operated by the Jet Propulsion
Laboratory, California Institute of Technology, under contract with the National Aeronautics
and Space Administration.  EB acknowledges support from the UK Science and Technology
Facilities Council [grant number ST/M001008/1]. FB acknowledges funding from the EU through
grant ERC-StG-257720.

\begin{deluxetable}{lccccc}
\tablenum{1} \tablecolumns{6} \tablewidth{390pt} \tablecaption{Global Properties
\label{tab-global}} \tablehead{ \colhead{Galaxy} & \colhead{$M(H_2)$} &
\colhead{$M(HI)_{\rm total}$} & \colhead{$M(HI)_{\rm CO field}$} & \colhead{$S_{\nu}(24\mu
m)$} &
\colhead{$S(H\alpha)$} \\
\colhead{} & \colhead{($M_\odot$)} & \colhead{($M_\odot$)} & \colhead{($M_\odot$)} &
\colhead{(mJy)} & \colhead{(erg cm$^{-2}$ s$^{-1}$)} } \startdata
IC 2163   &  $2.1\times10^9$  &  $4.8\times10^9$  &  $2.8\times10^9$  & $5.9\times10^2$  & $5.9\times10^{-13}$ \\
NGC 2207  &  $2.3\times10^9$  &  $2.2\times10^{10}$  &  $1.3\times10^{10}$  & $1.47\times10^3$  & $2.1\times10^{-12}$\\
\enddata
\end{deluxetable}

\begin{deluxetable}{lcccccccc}
\tablenum{2} \tabletypesize{\scriptsize} \tablecaption{Locations, Luminosities, and Surface
Densities\tablenotemark{a} \label{tab-location}} \tablewidth{0pt} \tablehead{ \colhead{ID}
& \colhead{R.A.}   & \colhead{Dec.}   & \colhead{$L$(24 \micron)\tablenotemark{b}} &
\colhead{$L(H\alpha)$\tablenotemark{c}}  & \colhead{SFR\tablenotemark{d}}  &
\colhead{$\Sigma_{\rm SFR}$\tablenotemark{e}} & \colhead{$\Sigma_{\rm
H_2}$\tablenotemark{e}} &
\colhead{$\Sigma_{\rm HI}$\tablenotemark{e}}   \\
&
\colhead{(J2000)} &
\colhead{(J2000)} &&&&&& \\
& \colhead{$06^h 16^m$}  & \colhead{$-21\degr$}    & &&&&&}
\startdata
  &   &    IC 2163&&&&&&\\
&&&&&&&&\\
A1\tablenotemark{g,h}
     &   30.612  & 22 47.20 &  19.2 &  0.195 & 0.0435 &  0.00748 &  13.6 & 15.6 \\
A2\tablenotemark{h}
     &   29.789  & 22 53.21 &  24.6 &  0.270 & 0.0568 &  0.00977 &  20.6 & 17.2 \\
A3 &   28.858  & 22 52.71 &  60.9 &  0.990 & 0.158   &  0.0272   &  35.0 &  9.8 \\
A4 &   27.892  & 22 51.21 &  75.9 &  1.16   & 0.193   &  0.0332   &  31.9 &  5.3\\
A5 &   27.133  & 22 43.71 &  93.1 &  2.20   & 0.280   &  0.0482   &  28.1 &  9.5\\
A6 &   26.345  & 22 37.21 &  53.7 &  2.17   & 0.211   &  0.0363   &   7.3 & 10.1 \\
A7\tablenotemark{h}
     &   30.755  & 22 28.20 &   5.7  &  0.101 & 0.0153 &  0.00263 &   2.6 & 15.2 \\
A8 &   28.930  & 22 35.71 &  73.3 &  0.539 & 0.155   &  0.0267   & 19.7 &  7.5  \\
A9 &   28.715  & 22 24.21 & 174.  &  0.727 & 0.336   &  0.0578   & 36.9 & 11.2 \\
A10\tablenotemark{h}
     & 28.783  & 22 09.21 & 0.15  &  0.086 & 0.0055 &  0.00095 &   0.5 & 12.6 \\
A11\tablenotemark{h}
     & 28.071  & 22 16.71 & 123.  &  0.586 & 0.242   &  0.0416   & 52.0 & 13.3 \\
A12 & 27.175  & 22 14.71 &  43.3 &  1.11   & 0.135   &  0.0232   & 33.5 & 10.2 \\
A13 & 26.320  & 22 12.71 &  52.2 &  1.55   & 0.174   &  0.0299   & 20.7 &   7.1 \\
A14 & 27.072  & 22 30.21 &  44.5 &  1.23   & 0.144   &  0.0248   & 15.9 &   7.2 \\
A15 & 27.999  & 22 32.21 &  21.6 &  0.984 & 0.0910 &  0.0157   & 12.6 &   4.2 \\
&&&&&&&&\\
 & &    NGC 2207 &&&&&&\\
&&&&&&&&\\
A16 &  25.135  & 22 30.21 &  23.5 &  0.875 & 0.0882  &  0.0162  &  2.0 & 20.9 \\
A17\tablenotemark{h}
       &  25.386  & 22 18.21 &  44.0 &  0.862 & 0.122    &  0.0225  & 14.7& 27.2 \\
A18 &  24.599  & 22 06.71 &  25.3 &  0.576 & 0.0748  &   0.0138 & 14.0& 17.5 \\
A19 &  24.670  & 21 49.21 &  47.3 &  1.13   & 0.143    &  0.0263  &   9.2&18.8  \\
A20 &  23.704  & 21 47.71 &  61.0 &  1.02   & 0.160    &  0.0294  & 14.7& 20.2 \\
A21 &  23.024  & 21 41.21 &  92.9 &  1.17   &  0.223   &  0.0410  &   7.6& 18.9 \\
A22 &  22.200  & 21 39.21 &  47.7 &  0.834 &  0.127   &  0.0234  &   5.5& 20.9 \\
A23 &  21.091  & 21 42.71 &   4.9  &  0.182 &  0.0183 &  0.0034  &  7.4 & 16.4 \\
A24 &  23.381  & 22 11.71 &  88.0 &  1.20   &  0.216   &  0.0398  & 12.9 &12.3 \\
A25 &  23.167  & 22 28.21 &  22.2 &  0.570 &  0.0692 &  0.0127  & 15.9 &  6.2 \\
A26 &  22.450  & 22 37.21 &    4.2 &  0.339 & 0.0259  & 0.0048   &   7.5 & 4.5 \\
A27 &  21.592  & 22 42.21 &  15.3 &  0.411 & 0.0487  & 0.0090   & 11.9 &10.2 \\
A28 &  20.589  & 22 42.21 &  15.2 &  0.513 &  0.0541  & 0.0100  & 12.9 & 15.4\\
A29 &  20.519  & 22 27.71 &  49.3 &  0.551 &  0.114   &  0.0210  &   9.2 &  8.8 \\
A30 &  19.551  & 22 39.71 &  51.5 &  0.736 &  0.128   &  0.0236  & 17.8 &15.7 \\
A31 &  18.477  & 22 33.21 &  24.7 &  0.516 & 0.0705  & 0.0130   &   7.1 & 21.4\\
A32 &  18.514  & 21 54.71 &  96.8 & 1.19    & 0.230    & 0.0423   &   6.3 & 18.2\\
A33\tablenotemark{h}
       &  17.225  & 21 54.71 &$<0.01$& 0.090 &0.0050   & 0.00091  &  0.8  & 27.5\\
A34 &  17.798  & 22 14.21 &  22.9 &  0.670 & 0.0759  & 0.0140    &  2.8 & 19.2\\
i\tablenotemark{f}
       &15.865   & 22 02.21 & 866.  & 2.39    & 1.606     & 0.296     & 14.9 & 19.9\\
A36\tablenotemark{h}
       &  16.580  & 22 23.21 &  40.5 & 0.972  &  0.123  &  0.0226  &  4.8 & 28.5 \\
A38\tablenotemark{h}
       &  17.439  & 22 43.71 &  40.7 & 1.07    &  0.128    &  0.0236  &  1.5 & 26.9 \\
A39 &  17.511  & 22 55.21 &  17.6 & 0.883  &  0.0786  &  0.0145  & 0.07& 20.0 \\
A40 &  21.377  & 23 10.71 &  44.5 & 0.872  &  0.124    &  0.0228  & 10.3 &12.0\\
A41 &  23.740  & 22 59.21 &    8.7 & 0.520  &  0.0433  &  0.0080  &  5.7 & 13.5 \\
A37 &  23.811  & 22 01.21 &  20.5 & 0.774  & 0.0776   &   0.0143 &  8.7 & 14.8 \\
A42 &  20.589  & 22 13.71 &    2.2 & 0.136 & 0.0112   &   0.0022  &  5.3 &  7.4 \\
A43 &  19.122  & 22 56.71 &    4.8 & 0.305 & 0.0249   &   0.0046  &  3.2 & 12.8 \\
A35 &  19.301  & 21 47.21 &  16.1 & 0.375 & 0.0481  &    0.0089  &  5.2 & 14.0 \\
\enddata
\tablenotetext{a} {The units are as follows: R.A. in seconds, Dec. in arcminutes and
arcseconds; $L$(24 \micron) and $L$(\Halpha) in $10^{40}$ erg s$^{-1}$; SFR in \msunyr;
$\Sigma_{\rm SFR}$ in \msunyrkpc; $\Sigma_{\rm H_2}$ and $\Sigma_{\rm HI}$ in \msunpc.}
\tablenotetext{b} {$L$(24\micron) is $\nu L_\nu$ at 24 $\mu {\rm m}$.} \tablenotetext{c}
{$L$(\Halpha) is corrected for Galactic foreground extinction using $A_{\rm V}/1.28$ =
0.186 mag. $L(H\alpha)$ for apertures A3, A4, A5, A6, A12, A13, A14, and A15 have also been
corrected for foreground extinction by the outer arm of NGC 2207 in front of IC 2163. The
latter correction (see text) amounts to an average of $0.85 \pm 0.43$ mag at $H\alpha$.}
\tablenotetext{d} {SFR(\msunyr) = $5.5 \times 10^{-42} [L(H\alpha) + 0.031 L(24\mu\;{\rm
m})]$ (erg s$^{-1})$. } \tablenotetext{e} {Surface densities have been corrected to face-on
using $i = 40\degr$ for IC 2163 and $i = 35\degr$ for NGC 2207.} \tablenotetext{f} {\em
Feature i} \tablenotetext{g} {Apertures are $14^{\prime\prime}$ in diameter.}
\tablenotetext{h} {In IC 2163, apertures A1 and A2 are on parts of the massive HI cloud I3,
A7 is on I2, and A10 and A11 are on parts of I5. In NGC 2207, apertures A17, A33, A36, and
A38 are on massive HI clouds N6, N2, N3, and N4, respectively.}
\end{deluxetable}

\begin{figure}
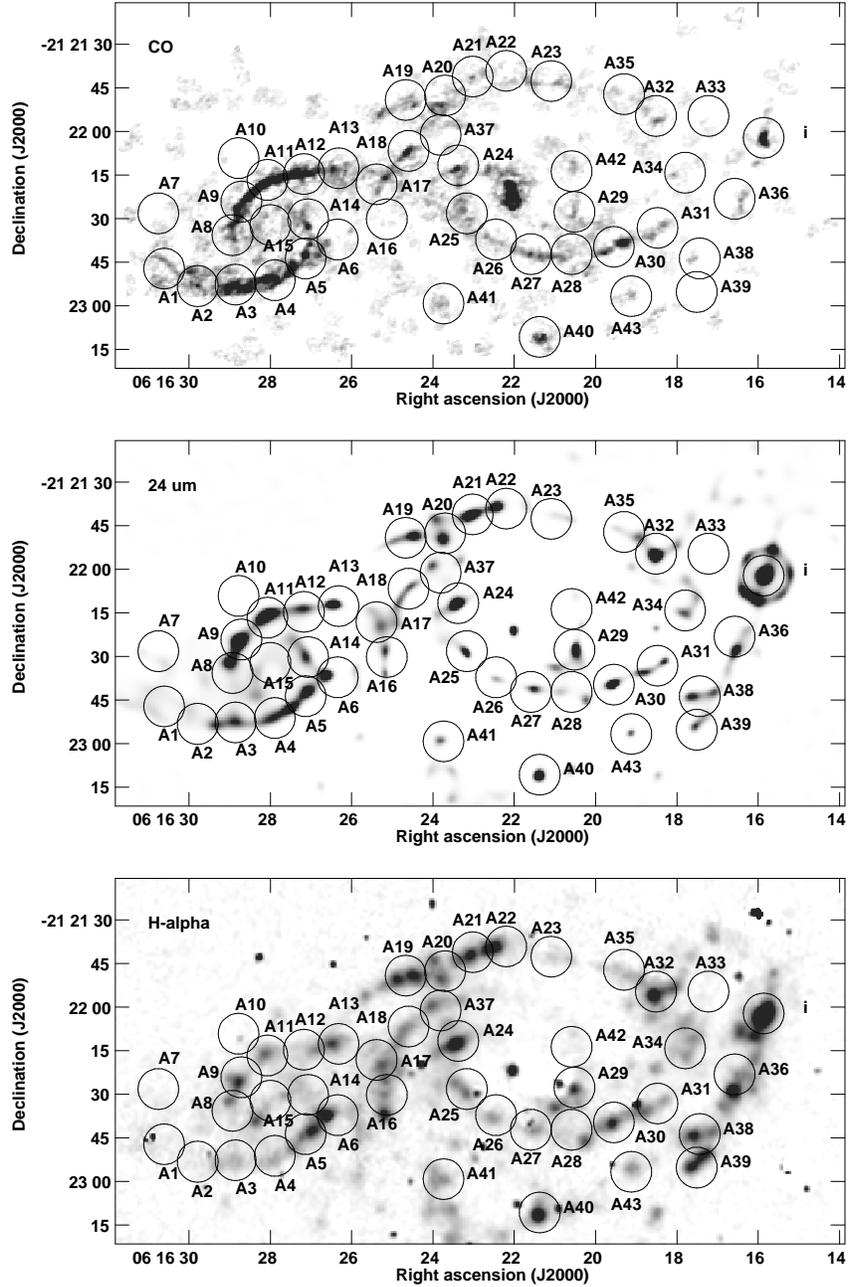

\epsscale{.7}
\plotone{f1a.eps}
\plotone{f1b.eps}
\plotone{f1c.eps}
\caption{(top) Integrated CO intensity (middle) HiRes 24  \micron\ flux density, and
(bottom) H$\alpha$ (uncorrected for extinction) with regions of star formation considered here.
The aperture size is $14^{\prime\prime} = 2.4$ kpc in diameter. \label{fig:CO}}
\end{figure}

\begin{figure}
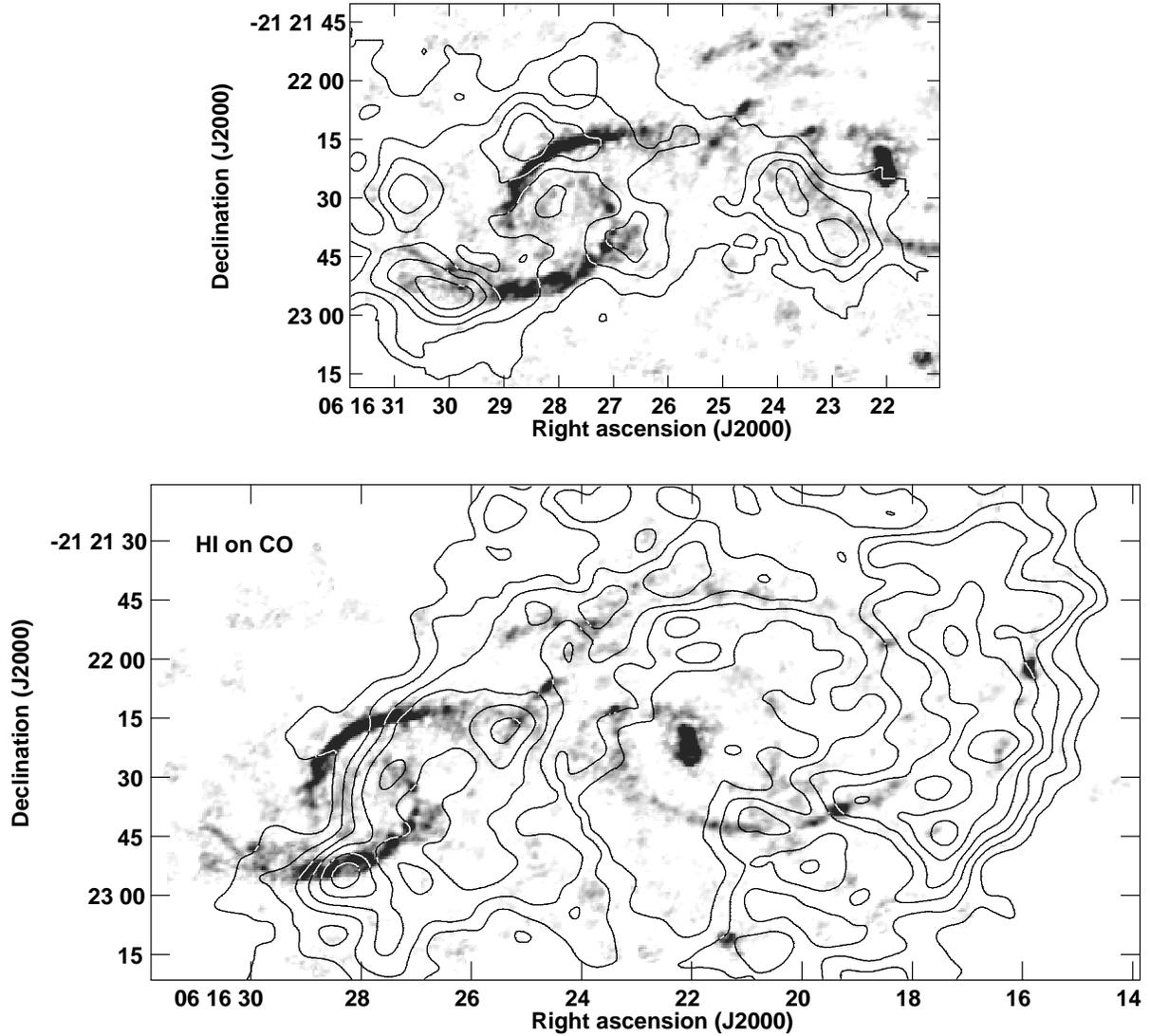

\epsscale{.66}
\plotone{f2a.eps}
\epsscale{1.}
\plotone{f2b.eps}
\caption{(top) HI column density contours superposed on CO for IC 2163. The contours are at line-of-sight
$N(HI) = 5$, 10, 15, 20, and $25\;M_\odot$ pc$^{-2}$. (bottom) HI on NGC 2207. The contours
are at 10, 15, 20, 25, 30, $35\;M_\odot$ pc$^{-2}$, with $5\;M_\odot$ pc$^{-2}$ omitted for
clarity. The HI contributions to each galaxy were determined on the basis of their kinematics.
\label{fig:HI}}
\end{figure}

\begin{figure}\epsscale{1.0} \plotone{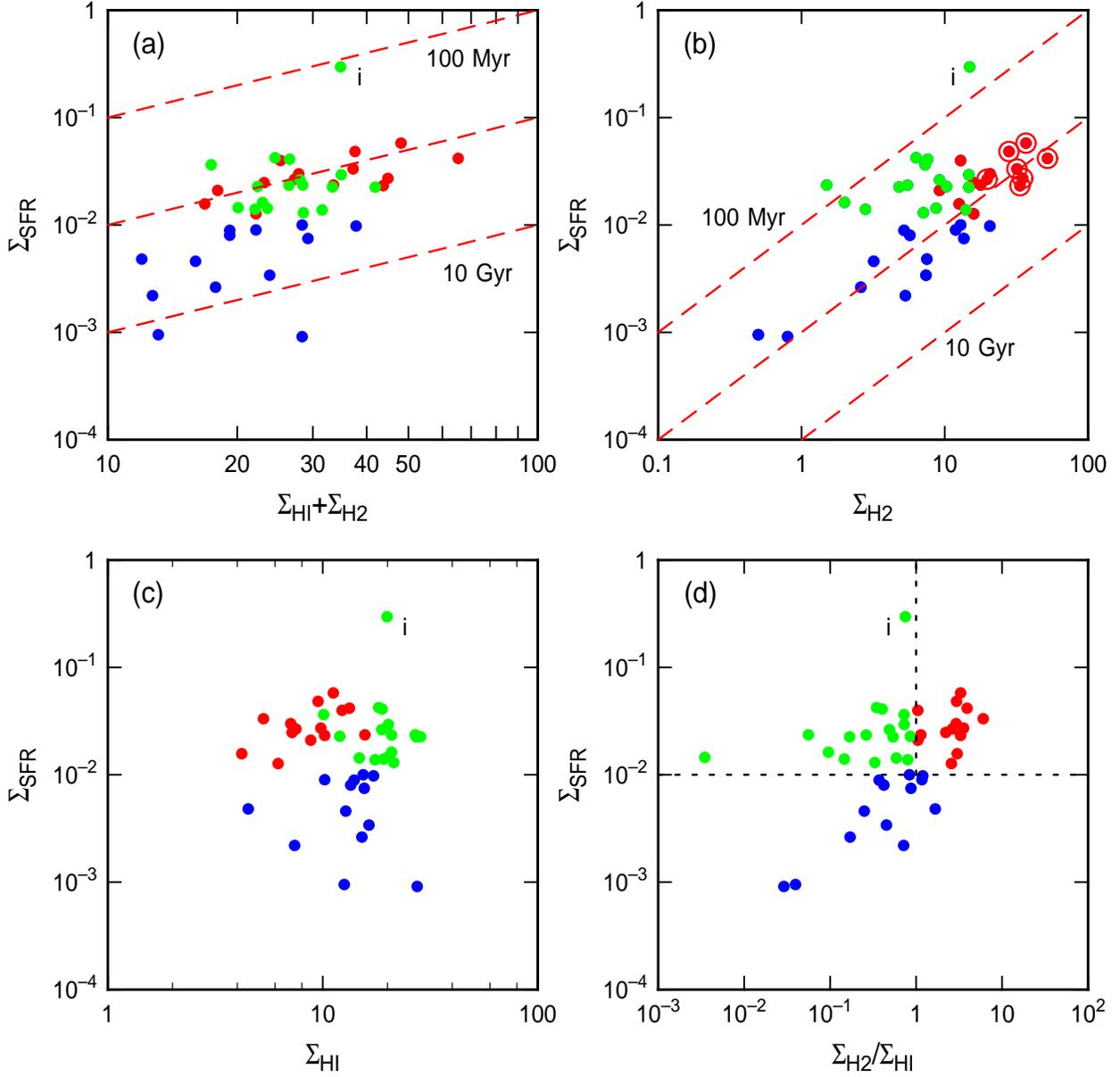}
\caption{The SFR surface density is plotted versus the gas surface density in various forms.
Color indicates regions with high SFR (green for low molecular fraction and red for
high molecular fraction) and low SFR (blue), as indicated in
panel (d). {\em Feature i} is labeled. The red circles in (b)
are for locations in the ocular rim of IC 2163. The red dashed lines in (a) and (b)
indicate gas consumption times. The dashed lines in (d) separate regions of high and low SFR
and high and low $\Sigma_{\rm H2}/\Sigma_{\rm HI}$ ratio.}
\label{fig:n2207_alma_ks1c}\end{figure}

\begin{figure}\epsscale{1.0} \plotone{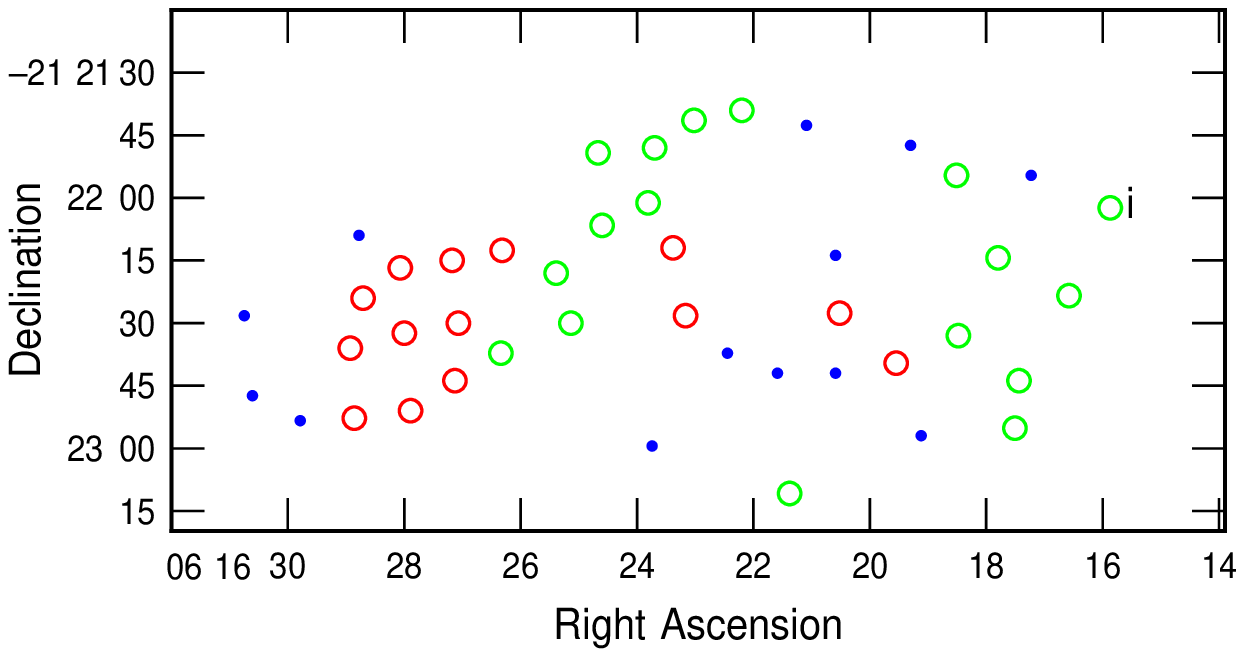}
\caption{Positions of the selected regions with color coding as in Figure 2. The green regions
have high SFRs yet are dominated by atomic gas. {\em Feature i} is on the right.}
\label{fig:n2207_alma_ks3}\end{figure}

\end{document}